\documentclass[twocolumn,aps,showpacs]{revtex4}
\usepackage{amsmath}
\usepackage{amssymb}
\usepackage{graphicx}
\usepackage{dcolumn}

\def\bphi{\mbox{\boldmath $\phi$}}
\def\bpsi{\mbox{\boldmath $\psi$}}
\def\bbpsi{\mbox{\boldmath $\Psi$}}

\begin{document}
\title{Effective Hamiltonian and dynamics of edge states in two-dimensional topological insulators 
under magnetic fields}
\author{O. E. Raichev}
\affiliation{Institute of Semiconductor Physics, National Academy of
Sciences of Ukraine, Prospekt Nauki 41, Kyiv 03028, Ukraine}
\email{raichev@isp.kiev.ua}




\begin{abstract}
The magnetic field opens a gap in the edge state spectrum of two-dimensional topological 
insulators thereby destroying protection of these states against backscattering. 
To relate properties of this gap to parameters of the system and to study dynamics 
of electrons in edge states in the presence of inhomogeneous potentials, the effective 
Hamiltonian theory is developed. Using this analytical theory, quantum-mechanical 
problems of edge-state electron transmission through potential steps and barriers, 
and of motion in constant electric field are considered. The influence of magnetic field 
on the resistance of two-dimensional topological insulators based on HgTe quantum wells 
is discussed together with comparison to experimental data.
\end{abstract}


\maketitle

\section{Introduction}

The two-dimensional (2D) topological insulators (also known as quantum spin Hall states) 
experimentally realized in HgTe-based quantum wells \cite{1,2,3,4,5} represent a unique state 
of matter characterized by the existence of a pair of helical counter-propagating 
one-dimensional (1D) conducting channels at the edges of 2D system. These edge states 
have gapless energy spectrum and are topologically protected due to time reversal 
symmetry, which means that electrons in these states move without backscattering if 
the inelastic scattering is neglected. When the Fermi energy stays in the bulk 
insulating gap, the edge states determine electron transport. The resistance 
quantization observed \cite{1} at low temperatures in the HgTe quantum well samples 
of macroscopic (1 $\mu$m) length and subsequently confirmed in nonlocal resistance 
measurements \cite{4} is a convincing proof for ballistic transport of electrons in the 
edge states.   

The problem of influence of magnetic fields on the transport properties of 2D topological 
insulators has appeared immediately after the discovery of these systems. Experimentally, 
it was found \cite{1,2,3} that the field perpendicular to the quantum well plane, $B_z$, 
considerably suppresses the conductance already at $B_z \sim 0.01$ T, while the in-plane 
magnetic field suppresses the conductance only when it is strong enough, of the order of 
$1$ T. It was also established that the conductance depends on the direction of the 
magnetic field in the quantum well plane \cite{3}. Later on, it was demonstrated \cite{6,7} 
that similar phenomena also take place in long samples, where the edge-state transport is not 
ballistic and the resistance is much larger than the resistance quantum. However, the effect 
of perpendicular magnetic field in these conditions is much smaller and manifests itself 
at $B_z \sim 1$ T, similar to the case of in-plane field. To describe the effect of the 
perpendicular field, several theoretical approaches have been proposed, including numerical 
simulation of transport based on a 2D disordered site model \cite{8}, as well as analytical 
consideration of the possibility of enhanced backscattering due to edge spectrum 
nonlinearity induced by the magnetic field \cite{9} and of interference effects at a disordered 
edge allowing for loops of the helical edge states \cite{10}. The approach of Ref. 8 gives 
an adequate description of the conductance suppression phenomenon under the assumption 
of strong disorder, when the amplitude of random potential perturbation exceeds the bulk 
gap energy. The behavior of the conductance under in-plane field has not been studied in 
detail, although numerical simulations \cite{8} also show a suppression of the conductance 
under the action of this field.    

A convenient analytical method for theoretical studies of edge-state transport in 2D 
topological insulators is based on the effective Hamiltonian theory, which assumes 
reduction of the 2D quantum-mechanical problem, which essentially requires numerical 
solution in the presence of disorder potential, to a 1D problem of electron motion 
along the edge. The effective $2 \times 2$ Hamiltonian for edge states \cite{11,12} has 
the form $\hbar v k\hat{\sigma}_z$, where $v$ is the edge-state velocity, 
$k$ is the operator of electron momentum along the edge and $\hat{\sigma}_z$ is the Pauli 
matrix. This Hamiltonian describes two branches of edge states with gapless energy spectrum. 
However, it does not take into account effects of magnetic field and potential perturbations. 
Meanwhile, it is well understood \cite{2,12} that external magnetic field (or the presence of 
magnetic impurities in the system) violates the time reversal symmetry and makes the edge 
states no longer gapless. The gap in the edge state spectrum means that the states with 
opposite spin projections mix with each other and, therefore, electron scattering between 
counter-propagating states becomes possible. Such a backscattering should reduce the 
conductance in a similar way as it takes place in constrictions \cite{13} or finite-width 
strips of 2D topological insulators, where the gap is formed because of the mixing of states 
from the opposite edges due to overlap of their wave functions \cite{14}. Therefore, inclusion 
of the gap opening effect of magnetic field into the Hamiltonian of edge states is a 
necessary step in developing the analytical theory of magnetotransport in 2D topological 
insulators. The most general form of such a $2 \times 2$ matrix Hamiltonian has been proposed 
in Ref. 12. It differs from the zero-field case by the presence of an additional term 
$\sum_{ij} a_{ij} B_i \hat{\sigma}_j$ which is linear in components of magnetic field 
$B_i$ (here $i$ and $j$ are Cartesian coordinate indices). For application of this 
Hamiltonian to HgTe quantum wells, the coefficients $a_{ij}$, which determine the effect 
of the magnetic field, have been calculated numerically \cite{12}. It is desirable, 
however, to have a more convenient representation of the effective Hamiltonian, where 
these coefficients are expressed through the known parameters of the HgTe quantum 
well system.

The basic goal of the research given in this paper is to develop the all-analytical 
effective Hamiltonian theory for 1D motion of electrons in the edge states in the presence 
of magnetic field, spin-orbit interaction, and potential perturbations. The key point 
of the forthcoming consideration is a transition from the two-subband 2D Hamiltonian 
of electrons in HgTe quantum wells \cite{2,5} to an effective 1D Hamiltonian for a couple 
of edge states. The influence of magnetic field is taken into account by using perturbation 
theory. The equation (7) derived in this way is similar to the massive Dirac equation 
describing 1D motion of spinless particles with a two-band energy spectrum. The gap 
between the bands in this spectrum considerably depends on the direction of the magnetic 
field and vanishes for a special orientation of the field. To show how the presence of 
the gap influences ballistic propagation of electrons in the edge states and how the 
backscattering probability depends on this field, a straightforward application of the 
theory to standard quantum-mechanical problems, such as transmission through potential 
barriers and motion in a uniform electric field, is carried out. Edge-state transport 
in the random potentials is also discussed. It is demonstrated that the gap in the edge 
state spectrum leads to a quadratic dependence of the resistance on the magnetic field 
strength. Some experimental magnetotransport data for 2D topological insulators, in 
particular, initial quadratic increase of the resistance as a function of the in-plane 
magnetic field and dependence of the resistance on the orientation of this field are 
explained within the proposed theory.

The paper is organized as follows. Section 2 is devoted to derivation of the equation 
for 1D motion in the edge states. Application of this equation to calculation of transmission 
and reflection probabilities, necessary for evaluation of the ballistic conductance via 
edge states, is done in Sec. 3. A discussion relating the results to experimental data 
on the conductance of 2D topological insulators based on HgTe quantum wells is presented 
in Sec. 4.

\section{Hamiltonian for edge states}

The effective $4\times4$ matrix Hamiltonian describing a 2D insulator phase in symmetric HgTe 
quantum well in the basis of two subbands (one of them is the interface-like subband 
formed as a result of hybridization of conduction-band states with light hole states 
and the other is the ground heavy hole subband, with quantization energies $\varepsilon_1$ 
and $\varepsilon_2$ respectively) was presented by Bernevig, Hughes, and Zhang \cite{5} 
and improved in subsequent works \cite{2,8,15,16} to account for spin-orbit coupling and 
magnetic fields. This Hamiltonian is written below in the approximation neglecting 
inessential small diagonal terms quadratic in electron momentum:
\begin{equation}
\hat{H}=\left( \begin{array}{cccc} 
U^{+}_{\bf r} +K_z B_z & A \hat{k}_+ & KB_- & -D \\
A \hat{k}_- & U^{-}_{\bf r}  & D & 0 \\ 
K B_+ & D & U^{+}_{\bf r} -K_z B_z & -A \hat{k}_-\\ 
-D & 0 & -A \hat{k}_+  & U^{-}_{\bf r}  \end{array} 
\right), 
\end{equation}
where $U_{\bf r}^{\pm}=\varphi_{\bf r} \pm \Delta_{\bf r}/2$, ${\bf r}=(x,y)$ is the 
coordinate in the quantum well plane, $\varphi_{\bf r}$ is the electrostatic potential, 
$\Delta_{\bf r} = \varepsilon_1 - \varepsilon_2$ is the gap energy whose possible variation in 
the plane is described as $\Delta_{\bf r}= \Delta + \delta_{\bf r}$, where $\Delta$ is the 
averaged gap, $\hat{k}_{\pm}=\hat{k}_{x} \pm i \hat{k}_{y} = -i \partial_x - 
e B_z y/\hbar c \pm \partial_y$, $e$ is the absolute value of electron charge, 
and $B_{\pm} = B_x \pm i B_y$. To describe the orbital effect of the perpendicular magnetic field 
$B_z$, the vector potential is chosen as $(-B_z y, 0)$. The coefficients $A$, $K$, $K_z$, 
and $D$ are parameters of the effective 2D Hamiltonian, they can be obtained directly by
applying the Kane Hamiltonian for calculation of eigenstates in the quantum well system 
(see Ref. 16 and references therein). 
In particular, $A$ describes coupling between the subbands at finite 2D momentum, $K$ and 
$K_z$ characterize effective Zeeman splitting of the interface-like subband due to in-plane 
and perpendicular magnetic fields (similar terms for the heavy hole subband are small enough 
to be neglected), and $D$ describes spin-orbit coupling due to bulk inversion asymmetry. As 
the upper left and lower right $2 \times 2$ blocks of the Hamiltonian (1) correspond to 
the states with opposite projections of spin, the terms $D$ and $KB_{\pm}$ are responsible 
for spin mixing. 

The Hamiltonian (1) can be presented in the form $\hat{H} = \hat{H}_0+\hat{H}_1$, where
\begin{equation}
\hat{H}_0=\left( \begin{array}{cccc} 
\Delta/2 & A \partial_y   & 0 & -D \\
-A \partial_y & - \Delta/2 & D &  0 \\ 
 0 & D & \Delta/2 &  A \partial_y \\ 
-D & 0 & -A \partial_y & -\Delta/2 \end{array} \right).
\end{equation}
Since the 2D system is supposed to be a topological insulator, the case of inverted 
subband ordering, $\Delta<0$, is considered below. The Hamiltonian $\hat{H}_0$ describes 
the states with zero component of momentum along $x$ axis in the absence of potential 
perturbations $\varphi_{\bf r}$, $\delta_{\bf r}$ and magnetic fields. 
Assume that there is a boundary at $y=0$ and the 2D system is at $y>0$. The 
eigenstate problem $(\hat{H}_0-\varepsilon) \bpsi(y)=0$ has a couple of edge-state 
solutions with wave functions evanescent from the boundary and energies inside 
the gap:
\begin{equation}
\left( \begin{array}{c} 1 \\
-u \\
-i \\
i u \end{array} \right) e^{-\kappa_+(\varepsilon) y} ,~~~ \left( \begin{array}{c} 1 \\
- u \\
i \\
-i u \end{array} \right) e^{-\kappa_-(\varepsilon) y},
\end{equation}
where $u=\sqrt{(|\Delta|/2+\varepsilon)/(|\Delta|/2-\varepsilon)}$ and 
$A \kappa_{\pm}(\varepsilon)=\sqrt{(\Delta/2)^2-\varepsilon^2} \pm iD$. To complete 
the eigenstate problem, the Hamiltonian $\hat{H}_0$ should be supplemented with 
boundary conditions which allow one to find proper orthogonal combinations of 
the wave functions (3) and corresponding eigenvalues of energy. Without specifying 
these conditions, let us apply a simplifying assumption that the particle-hole 
symmetry of the Hamiltonian $\hat{H}_0$ is retained in the presence of the boundary. 
This case is realized, for example, if the normal insulator state at $y<0$ is described 
by a Hamiltonian of the same form as Eq. (2) but with large positive $\Delta$; then the 
boundary conditions take the form \cite{9} $\psi_1(0)=-\psi_2(0)$ and $\psi_3(0)=
-\psi_4(0)$, where $\psi_i$ are the components of the columnar wave function $\bpsi$. 
Under this assumption, the eigenstates are double degenerate, with energy $\varepsilon=0$, 
so $u=1$. The functions (3) are orthogonal to each other under these conditions. It is 
convenient to form the basis of eigenstates by using normalized orthogonal combinations 
of these functions, $\bbpsi_1(y)$ and $\bbpsi_2(y)$, according to
\begin{equation}
\bbpsi_n(y) = C_{n} \left( \begin{array}{c} 1 \\
-1 \\
-i \\
i \end{array} \right) e^{-\kappa_+ y} + C^*_{n} \left( \begin{array}{c} 1 \\
-1 \\
i \\
-i \end{array} \right) e^{-\kappa_- y},
\end{equation}
where $n=1,2$ numbers the edge states, 
\begin{equation}
C_{1}= \frac{\sqrt{\kappa}}{2} \left(\frac{\kappa_+}{\kappa_-} \right)^{1/4},~ C_{2}=i C_{1},   
\end{equation}
and
\begin{equation}
\kappa_{\pm}=\kappa(1 \pm i \mu),~ \kappa=\frac{|\Delta|}{2A},~ \mu=\frac{2D}{|\Delta|}.  
\end{equation}
The quantity $\kappa$ is the inverse length of decay of edge-state wave function from the 
boundary. The dimensionless parameter $\mu$ characterizes strength of spin-orbit coupling.

Applying this basis, i.e. searching for the solution of the eigenstate problem with 
Hamiltonian (1) in the form $\phi_1(x) \bbpsi_1(y)+ \phi_2(x) \bbpsi_2(y)$, one obtains 
a $2 \times 2$ matrix equation describing 1D motion in the edge states: 
\begin{eqnarray}
(\hat{h} -\varepsilon) \bphi(x)=0,~~~~~~~~~~~~~~~~~~ \\
\hat{h}=\left( \begin{array}{cc} \varphi_x +i \hbar v \partial_x  + vp_0 & g_{\bf B} \\
g^*_{\bf B} & \varphi_x - i \hbar v \partial_x  - vp_0 
\end{array} \right), \nonumber
\end{eqnarray}
where $\bphi=(\phi_1,\phi_2)^T$ is the columnar wave function of the edge states and 
$\varphi_x= 2 \kappa \int_0^{\infty} dy \varphi_{\bf r} e^{-2 \kappa y}$ 
is the electrostatic potential acting at the edge. The parameters entering the diagonal 
part of $\hat{h}$ are the edge state velocity $v$ and momentum displacement $p_0$ 
defined as  
\begin{equation}
\hbar v =\frac{A}{\sqrt{1+\mu^2}},~p_0=\frac{eB_{z}}{2 \kappa c (1+\mu^2)} 
+ \frac{K_z B_z}{2 v \sqrt{1+\mu^2}}.
\end{equation}
The complex energy entering the non-diagonal part is linear in magnetic field: 
\begin{equation}
g_{\bf B}= \frac{K B_x}{2 \sqrt{1+\mu^2}} - i \frac{K B_y}{2} + \frac{\mu e v B_{z}}{2 \kappa c (1+\mu^2)}.
\end{equation}
If the magnetic field is directed in the $(xz)$ plane, $g_{\bf B}$ is real. The 
contribution of the magnetic field into the effective Hamiltonian $\hat{h}$ has the
general form given in Ref. 12. However, in contrast to Ref. 12, all the parameters 
are now explicitly related to the quantities entering the Hamiltonian of 2D electrons 
in HgTe quantum wells, Eq. (1). 

The usage of truncated basis $\bbpsi_1(y)$ and $\bbpsi_2(y)$ (which is actually the 
key approximation in derivation of any effective Hamiltonian) provides limitations 
for application of Eq. (7). This equation cannot be used in the region of energies 
beyond the 2D gap, $|\varepsilon - \varphi_x| > |\Delta|/2$, where the extended 2D 
states exist together with the edge states and can mix with them. Such mixing occurs 
even at $|\varepsilon - \varphi_x| < |\Delta|/2$ if the fluctuations of the gap, 
$\delta_{\bf r}$, are strong. For this reason, $|\delta_{\bf r}|$ has to be small 
compared to $|\Delta|$. In fact, the gap fluctuations do not enter Eq. (7), because 
the latter is obtained in the first order of perturbation theory in the Hamiltonian 
$\hat{H}_1$. Calculation of the second-order contributions leads to additional 
terms associated with $\delta_{\bf r}$, such terms can be identified as 
coordinate-dependent corrections to the quantities $\kappa$ and $\mu$ entering 
$v$, $p_0$, and $g_{\bf B}$ in Eq. (7). Since these corrections are assumed to be 
small, Eq. (7) is justified. In spite of the limitations described, the expression 
(8) for the edge state velocity $v$ is exact (valid for arbitrary energy) at ${\bf B}=0$. 

Equation (7) is similar to one-dimensional Dirac equation for massive spinless particles. 
Consider the main properties of this equation. First of all, the terms $\pm v p_0$ entering 
the diagonal part of $\hat{h}$ describe a shift of electron spectrum in momentum space, due 
to the perpendicular magnetic field. These terms have no effect on edge state propagation \cite{9}
and can be removed from Eq. (7) by a substitution $\bphi(x) \rightarrow \bphi(x) 
\exp(i p_0 x/\hbar)$ equivalent to a gauge transformation. For this reason, the terms 
$\pm v p_0$ in $\hat{h}$ are omitted below. 

The free motion in edge states, when $\varphi_x=0$, is characterized by constant momentum 
$p$ and described by the energy spectrum  
\begin{equation}
\varepsilon_p = \pm \sqrt{ v^2 p^2 + |g_{\bf B}|^2},
\end{equation}
which has a gap equal to $2|g_{\bf B}|$. Both the in-plane field and the 
perpendicular field contribute to the gap. However, the contribution of the 
perpendicular field appears only due to orbital effect of this field and in 
the presence of spin-orbit coupling \cite{2}, when $\mu \neq 0$, while the contribution 
of the in-plane field does not require this coupling \cite{16}. In the case of tilted 
field with $B_y=0$ and $B_z/B_x=-\kappa c K \sqrt{1+\mu^2}/e v \mu$ the gap disappears. 
Making estimates for HgTe quantum well of width 7.3 nm ($|\Delta| \simeq 20$ meV) and 
assuming $D \simeq 1$ meV, one finds that the ratio $B_z/B_x$ corresponding to zero 
gap is of the order of unity. The origin of gap disappearance can be explained as a 
mutual compensation of the orbital and Zeeman effects of the magnetic field, which are 
caused by $B_z$ and $B_x$, respectively. 
  
If the gap is equal to zero, Eq. (7) has two exact solutions:
\begin{eqnarray}
\bphi^{(1)}(x)=\left( \begin{array}{c} e^{-i \Phi_x} \\
0 \end{array} \right),~\bphi^{(2)}(x) = \left( \begin{array}{c} 0 \\
e^{i \Phi_x} \end{array} \right) , \\
\Phi_x= \frac{1}{\hbar v} \int^x dx' (\varepsilon- \varphi_{x'}). \nonumber
\end{eqnarray}
Since the probability current for the states of Eq. (7) is equal to 
$v(|\phi_2(x)|^2-|\phi_1(x)|^2)$, these solutions describe constant probability currents 
equal to $-v$ (left-moving) and $v$ (right-moving), respectively. Regardless to the form 
of the potential $\varphi_{x}$, there is no backscattering. Thus, the approach presented 
above relates the topological protection of edge states to a basic property of massless 
1D Dirac model. This property is responsible for ideal transmission of electrons through 
potential barriers (Klein paradox \cite{17}), observed in graphene at normal incidence 
\cite{18,19}. In the case of nonzero gap, the transmission depends on details of the 
potential, as illustrated below in some examples of quantum-mechanical problems.

\section{Propagation of gapped edge states}

Consider propagation of electrons in the edge states described by Eq. (7) through finite-size 
regions with inhomogeneous electrostatic potential. Since the methods of solutions of 
similar quantum-mechanical problems for Dirac model are well known \cite{17}, and the 
subject is under increasing attention in connection with studies of graphene \cite{18}, 
the details of calculations leading to the results presented below are omitted. 
The electric current carried by a single edge through an inhomogeneous region 
is given by $j=(e/h) \int d \varepsilon T_{\varepsilon}[f_L(\varepsilon)-f_R(\varepsilon)]$, 
where $f_L$ ($f_R$) are the energy distribution functions of electrons in the edge states on 
the left (right) sides of the potential perturbation and $T_{\varepsilon}$ is the transmission 
probability related with the reflection probability $R_{\varepsilon}$ in the usual way, $T_{\varepsilon}+R_{\varepsilon}=1$. The simplest important cases are the following. 

\begin{figure}[ht]
\begin{center}\leavevmode
\includegraphics[width=8.6cm]{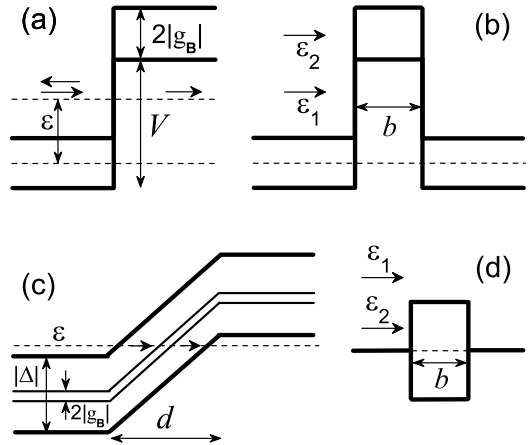}
\label{Fig. 1} \caption{Transmission through a potential step (a), a potential barrier (b),
interband transmission in a homogeneous electric field created in a Zener diode 
structure (c), and transmission through a magnetic barrier defined by a local gap 
at $|x|<b/2$ (d). Solid lines plotted at $\varphi_x \pm |g_{\bf B}|$ denote the potential 
energies of the conduction and valence bands of edge-state 1D electrons; in the part (c) 
the band edges of 2D electrons, at $\varphi_x \pm |\Delta|/2$, are also plotted.}
\end{center}
\end{figure}

{\em Potential step}. In the case of the potential step ($\varphi_x=0$ at $x<0$ and $\varphi_x=V$ 
at $x>0$) shown in Fig. 1 (a) the transmission of electron with energy $\varepsilon$ is nonzero 
if $|V-\varepsilon|>|g_{\bf B}|$. Introducing $\lambda_1=|g_{\bf B}|/\varepsilon$ and 
$\lambda_2=|g_{\bf B}|/(V-\varepsilon)$ one gets
\begin{equation}
R_{\varepsilon}=\frac{|\eta_1+\eta_2|^2}{|1+\eta_1\eta_2|^2},~~\eta_i=\frac{\lambda_i}{\sqrt{1-\lambda_i^2}+1}. 
\end{equation}
This expression is formally valid also at $|V-\varepsilon|<|g_{\bf B}|$, when $\eta_2$ 
is a complex quantity; in this case the reflection is perfect, $R_{\varepsilon}=1$. The expression 
(12) is considerably simplified for a special condition $\varepsilon=V/2$ (symmetric $n-p$ 
junction, as shown in Fig. 1 (a)): $R_{\varepsilon}=4|g_{\bf B}|^2/V^2$. If 
the gap is small so that $|\lambda_i| \ll 1$, the reflection probability is small 
and quadratic in ${\bf B}$: $R_{\varepsilon} \simeq |g_{\bf B}|^2 V^2/[2\varepsilon(V-\varepsilon)]^2$.

{\em Potential barrier}. The case $\varphi_x=0$ at $|x|>b/2$ and $\varphi_x=V$ at $|x|<b/2$ 
represents the potential barrier shown in Fig. 1 (b). The reflection coefficient is 
\begin{equation}
R_{\varepsilon}=\frac{\sin^2(k b)}{\sin^2(k b)+Q},~~Q=\frac{(1-\lambda_1^2)(1-\lambda_2^2)}{
(\lambda_1+\lambda_2)^2}, 
\end{equation}
where $k=\sqrt{(V-\varepsilon)^2-|g_{\bf B}|^2}/\hbar v$ is the wavenumber of electron inside the 
barrier. If $|V-\varepsilon|>|g_{\bf B}|$ ($\varepsilon=\varepsilon_1$ in Fig. 1 (b)), electrons can 
propagate in the barrier region, $k$ is real, and $R_{\varepsilon}$ demonstrates interference 
oscillations and goes to zero each time when $k b/\pi$ is integer. Since $k$ is controlled by 
the magnetic field, the oscillations can be induced by sweeping ${\bf B}$. If $|V-\varepsilon|< 
|g_{\bf B}|$ ($\varepsilon=\varepsilon_2$ in Fig. 1 (b)), electrons cannot propagate in 
the barrier region. In this case $k$ is imaginary, $k=i|k|$, $\sin^2(k b)=-\sinh^2(|k| b)$, 
and transmission through sufficiently wide barriers is exponentially small, 
$T_{\varepsilon} \propto \exp(-2 |k| b)$. If $|\lambda_i| \ll 1$, 
the reflection is small, $R_{\varepsilon} \simeq |g_{\bf B}|^2 V^2 \sin^2(k b)/[\varepsilon(V-\varepsilon)]^2$.

{\em Homogeneous electric field}. The motion of a Dirac electron in the potential $\varphi_x= Fx$
assumes transition of this electron between conduction and valence bands in the region near 
the point $\varepsilon=\varphi_x$. For massless Dirac Hamiltonian this transition takes place 
with unity probability. For massive Dirac Hamiltonian the electron undergoes interband (Zener) 
tunneling, which is formally described within an exactly solvable problem of tunneling 
through a parabolic barrier \cite{20}. The energy-independent transmission 
probability is given by the expression
\begin{equation}
T= \exp \left(-\pi |g_{\bf B}|^2/\hbar F v \right).
\end{equation}
A similar expression describes Zener tunneling in graphene at nonzero angle of incidence \cite{21}.
If the magnetic field is weak so that $|g_{\bf B}|^2 \ll \hbar F v/\pi$, one gets a small reflection 
probability quadratic in magnetic field, $R=\pi |g_{\bf B}|^2/\hbar Fv$. The 
homogeneous electric field can be created, for example, in planar $n-p$ junctions \cite{22} formed in 
HgTe quantum well structures with split gates, Fig. 1 (c). In the case of normal insulator the 
conductance of such a junction is proportional to $L_y \exp(-\pi \Delta^2/4 F A)$, where $L_y$ is 
the sample width. In the case of topological insulator, the edge state transport causes an additional 
contribution to conductance \cite{22}, which is equal to $2e^2/h$ in the absence of magnetic fields 
and can dominate over the normal (bulk) contribution as the latter is exponentially small at 
$\Delta^2 \gg F A$. If the magnetic field is present, the edge-state contribution is multiplied 
by a factor of $T$ from Eq. (14). Since $F \simeq |\Delta|/d$, where $d$ is the length of the 
inhomogeneous region, the factor in the exponent of Eq. (14) is estimated as $-2 \pi \kappa 
d |g_{\bf B}/\Delta|^2$. For $d$ of micrometer scale, one has $\kappa d \gg 1$, 
because $\kappa^{-1} \sim 30$ nm. This means that, despite of the assumed smallness of the 
ratio $|g_{\bf B}|/|\Delta|$, an exponentially small edge-state transmission in planar $n-p$ 
junctions is feasible in the presence of a magnetic field. Therefore, application of a magnetic 
field can be used for efficient control of the current through Zener diode in the topological 
insulator state. 

Notice that Eqs. (12)-(14) coincide with corresponding results for one-dimensional Dirac 
model \cite{17}, where $|g_{\bf B}|$ plays the role of a "mass parameter". The reason for 
this is the following: although the effective Hamiltonian ${\hat h}$ is formally different
from one-dimensional Dirac Hamiltonian, it can be reduced to the latter by a unitary 
transformation. 

{\em Magnetic barrier}. The case of inhomogeneous magnetic field is described by Eq. (7) 
with $x$-dependent $g_{\bf B}$. Consider a rectangular magnetic barrier, implying 
that the field is absent at $|x|>b/2$ and finite at $|x|<b/2$. This leads to the potential 
configuration where the gap exists locally at $|x|<b/2$, as shown in Fig. 1 (d). The reflection 
probability is
\begin{equation}
R_{\varepsilon}=\left(1+ \frac{\hbar^2 v^2 k^2}{|g_{\bf B}|^2 \sin^2(k b)} \right)^{-1},
\end{equation}
where $k=\sqrt{\varepsilon^2-|g_{\bf B}|^2}/\hbar v$. Similar to the case of potential barrier,
electrons either can or cannot propagate in the barrier region: $|\varepsilon|>|g_{\bf B}|$, real 
$k$ or $|\varepsilon|<|g_{\bf B}|$, imaginary $k$, respectively. In the first situation the 
transmission probability shows interference oscillations, while in the second one the transmission 
can be exponentially small. The model of magnetic barriers is suitable for description of electron 
scattering by magnetic impurities in the absence of magnetic field. Since the spin-dependent 
interaction of electron with such an impurity occurs at a short range, it is natural to assume 
small $b$ in Eq. (15) for such applications. If the product $k b$ is also small, Eq. (15) 
describes a small energy-independent reflection probability $R \simeq |g_{\bf B}|^2 b^2/\hbar^2 v^2$. 

The validity of the results of this section is determined by the range of applicability 
of the effective Hamiltonian (7). In particular, an estimate of the second-order 
perturbation contributions to $\hat{h}$ allows one to conclude that the results are 
applicable in the range of magnetic fields where the energies $|KB|$ and $|K_zB_z|$ 
are small in comparison to $|\Delta|$. For HgTe wells both $K$ and $K_z$ are estimated 
as 1 meV/T, so the results are justified within 1 Tesla range provided that 
$|\Delta| > 10$ meV.   

\section{Discussion}

The opening of the gap in the edge-state spectrum of 2D topological insulators under the 
action of magnetic field is the fundamental consequence of violation of the time reversal 
symmetry. A one-dimensional quantum-mechanical equation derived in this paper describes 
motion of electrons in the edge states in these conditions. The contribution of the in-plane 
field to the gap energy $2|g_{\bf B}|$ arises because of Zeeman interaction, while the 
contribution of the perpendicular field exists purely due to orbital effect of magnetic 
field in the presence of spin-orbit coupling. The gap is sensitive not only to the 
strength of the magnetic field but also to the direction of this field, and vanishes 
at a special field orientation. 

\begin{figure}[ht]
\begin{center}\leavevmode
\includegraphics[width=8.6cm]{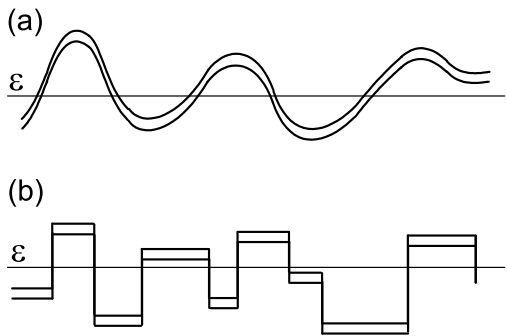}
\label{Fig. 2} \caption{The energies of the band extrema for edge-state 1D electrons, given 
by the expression $\varphi_x \pm |g_{\bf B}|$, in the cases of smooth (a), and sharp (b) 
random potentials. Electrons with energy $\varepsilon$ experience multiple Zener tunneling 
(a) or multiple transmission across potential steps (b). In both cases the reflection 
probability is quadratic in magnetic field.}
\end{center}
\end{figure}

When the gap is present, the scattering of electrons between counter-propagating edge channels 
can be responsible for suppression of the ballistic conductance in HgTe quantum well samples. 
Let us consider a model of transport in a disordered HgTe quantum well near the charge neutrality 
point, when Fermi energy is in the middle of the gap between 2D subbands. Since the magnetic field 
creates a small gap $2|g_{\bf B}|$, an edge-state electron moving in a spatially smooth 
random potential $\varphi_x$, see Fig. 2 (a), experiences multiple Zener tunneling events 
for which the averaged potential energy tilt $F$ is estimated as $2 w/l_c$, where $w$ is 
the mean amplitude of the potential variations ($w$ is assumed to be much larger than 
$|g_{\bf B}|$) and $l_c$ is the mean distance between adjacent local minima and maxima 
of the random potential. Each tunneling event occurs with small reflection probability, 
but the total number of such transitions is large and estimated as $N=L/l_c$, where $L$ 
is the sample length. Neglecting interference effects, one may roughly estimate the total 
reflection probability as $NR$, with $R=1-T$ taken from Eq. (14). Then, the relative increase 
in resistance is expected to be of the order of $\pi |g_{\bf B}|^2 L/2 \hbar v w$. If the 
magnetic field is perpendicular to the quantum well plane, this dimensionless quantity is estimated 
as $2 \pi A^3 D^2 L/|\Delta|^4 \ell^4 w$, where $\ell=\sqrt{\hbar c/e|B_z|}$ is the magnetic 
length. Using $A=360$ meV nm, $|\Delta|=20$ meV, $D=1$ meV, and assuming $w=3$ meV and 
$L=2$ $\mu$m, the relative increase in resistance is given by $2.5 [B_z ({\rm T})]^2$. If 
the random potential is sharp, as shown in Fig. 2 (b), the electron is transmitted across a 
number of potential steps, and the relative increase in resistance is estimated as 
$|g_{\bf B}|^2 L/w^2 l_c$, where $l_c$ is now the average distance between potential 
steps. Assuming $l_c \simeq 10$ nm, one gets a stronger effect of the magnetic field on 
the resistance compared to the case of smooth random potential. However, the model of 
smooth potential seems to be a more realistic one.

Therefore, according to the theory described in this paper, the increase in resistance is 
proportional to the square of the magnetic field and becomes essential in the fields of 
the order 1 T. Since the contributions of the in-plane field and the perpendicular field 
to the gap energy are comparable (see Sec. 2), the effect of both fields on backscattering 
should be similar. In agreement with this theory, the numerical simulations of transport 
on a 2D disordered site model \cite{8} show a weak, quadratic in magnetic field increase 
in resistance in the case of weak disorder. For strong disorder, when the potential amplitude 
exceeds the bulk gap energy, the simulation gives a completely different result: a strong, 
linear in $B_z$ increase in resistance and much weaker effect of the in-plane field. Since 
it is the behavior actually observed experimentally, one may conclude that the strong 
disorder is a necessary precondition for the dramatic influence of the perpendicular 
magnetic field on ballistic resistance and that the strong disorder is present in the 
samples investigated. The case of strong disorder, when transitions of electrons between 
the edge and bulk states are possible, is beyond the range of validity of the effective 
Hamiltonian theory.

On the other hand, the theory can explain the observed \cite{3,6} increase in resistance 
under the in-plane magnetic field of the order 1 T. This increase apparently begins with 
a quadratic field dependence and is sensitive to the gate voltage controlling 
position of the Fermi energy in the bulk gap \cite{6}. Moreover, the dependence of the 
resistance on orientation of the in-plane field shows (Ref. 3, p. 68) that the field 
perpendicular to the current leads to a stronger effect. This is in agreement with 
the theory demonstrating that the field $B_y$ creates a larger gap than the field $B_x$, 
see Eq. (9). The difference is controlled by the parameter $\mu=2D/|\Delta|$, which is 
not small in the case when the bulk gap $|\Delta|$ is of the order of 1 meV. Indeed, 
a considerable dependence of the resistance on orientation of the field has been 
observed \cite{3} for a sample with a small bulk gap. Numerical simulations \cite{8}, 
carried out for $\mu \ll 1$, show a negligible orientation dependence, in 
agreement with the theory. 

In summary, the parameters of the effective Hamiltonian for edge states in 
two-dimensional topological insulators based on HgTe quantum wells are calculated 
analytically. The effect of magnetic fields on quantum-mechanical properties of 
these states in the presence of potential perturbations is studied. A feasibility 
of broad control of ballistic transport of edge-state electrons by magnetic fields 
is demonstrated. The results of the effective Hamiltonian theory are compared to 
both experimental findings and numerical simulation data on magnetoresistance of 
HgTe quantum wells. The theory explains the observed behavior of resistance under 
the in-plane magnetic field, but does not explain a strong linear increase in 
resistance as a response to perpendicular magnetic field. This increase is likely 
caused by the transitions of electrons between edge and bulk states, which are not 
accounted by the effective Hamiltonian theory and are possible in the presence of 
high-amplitude disorder potential.

\end{document}